\documentclass[fleqn,usenatbib]{mnras}
\usepackage{graphicx}
\usepackage{natbib}
\usepackage{amssymb}
\usepackage{amsmath}
\usepackage{array,booktabs}
\usepackage{mathptmx}
\usepackage{booktabs}
\usepackage{hyperref}
\usepackage{float}
\newcommand{\msun}{\,{\rm M_{\odot}}}

\newcommand{\vp}{\,{v_{\rm ph}}}

\newcommand{\appropto}{\mathrel{\vcenter{
  \offinterlineskip\halign{\hfil$##$\cr
    \propto\cr\noalign{\kern2pt}\sim\cr\noalign{\kern-2pt}}}}}
\usepackage{xcolor}
\usepackage[normalem]{ulem} 

\title[Shocked jets in CCSNe can power FBOTs]{Shocked jets in CCSNe can power the zoo of fast blue optical transients}
\author[Gottlieb, Tchekhovskoy \& Margutti]{
	Ore Gottlieb$^{1}$\thanks{ore@northwestern.edu},
	Alexander Tchekhovskoy$^{1}$,
	Raffaella Margutti$^{2}$
	\\
	$^{1}${Center for Interdisciplinary Exploration \& Research in Astrophysics (CIERA), Physics \& Astronomy, Northwestern University, Evanston, IL 60201, USA}\\
	$^{2}${Department of Astronomy and Astrophysics, University of California, Berkeley, CA 94720, USA}\\
}
\pubyear{2022}
\begin{document}
	\label{firstpage}
	\pagerange{\pageref{firstpage}--\pageref{lastpage}}
	\maketitle	
	\begin{abstract}

    Evidence is mounting that recent multiwavelength detections of fast blue optical transients (FBOTs) in star-forming galaxies comprise a new class of transients, whose origin is yet to be understood. We show that hydrogen-rich collapsing stars that launch relativistic jets near the central engine can naturally explain the entire set of FBOT observables. The jet-star interaction forms a mildly-relativistic shocked jet (inner cocoon) component, which powers cooling emission that dominates the high velocity optical signal during the first few weeks, with a typical energy of $ \sim 10^{50}-10^{51} $ erg. During this time, the cocoon radial energy distribution implies that the optical lightcurve exhibits a fast decay of $ L \appropto t^{-2} $. After a few weeks, when the velocity of the emitting shell is $ \sim 0.01 $ c, the cocoon becomes transparent, and the cooling envelope governs the emission. The interaction between the cocoon and the dense circumstellar winds generates synchrotron self-absorbed emission in the radio bands, featuring a steady rise on a month timescale. After a few months the relativistic outflow decelerates, enters the observer's line of sight, and powers the peak of the radio lightcurve, which rapidly decays thereafter. The jet (and the inner cocoon) become optically thin to X-rays $ \sim $ day after the collapse, allowing X-ray photons to diffuse from the central engine that launched the jet to the observer. Cocoon cooling emission is expected at higher volumetric rates than gamma-ray bursts (GRBs) by a factor of a few, similar to FBOTs. We rule out uncollimated outflows, however both GRB jets and failed collimated jets are compatible with all observables.
    
    \end{abstract}

	
	\section{introduction}\label{sec:introduction}
	
	The increasing number of high cadence optical surveys over the past years have brought to light a variety of rapidly evolving supernova-like transients \citep[e.g.,][]{Poznanski2010,Drout2013,Drout2014,Arcavi2016,Shivvers2016,Tanaka2016,Whitesides2017,Prentice2018,Pursiainen2018,Rest2018,Tampo2020,Karamehmetoglu2021,Perley2021,Yao2021}. There is growing evidence that some of these are in some tension with traditional supernova (SN) models, implying at a new class of transients, characterized by rapid and luminous optical emission, known as Fast Blue Optical Transients \citep[FBOTs;][]{Drout2014,Margutti2019}. The rate of these events, which primarily take place in star-forming dwarf galaxies, is still under debate, but recent studies estimate it to be on order of $ \sim 1\% $ of core-collapse SNe \citep[CCSNe;][]{Tanaka2016,Ho2020,Coppejans2020}.
	In addition to optical emission, five FBOTs-AT2018lug, AT2018cow, CSS161010, AT2020xnd and AT2020mrf have shown evidence of radio signals, and the last four also shined in X-rays \citep{Coppejans2018,Coppejans2020,RiveraSandoval2018,Margutti2019,Ho2019,Ho2020,Ho2021,Yao2021,Bright2021}.
	
	To date, all radio signals associated with FBOTs have shown a consistent behavior with synchrotron self-absorption (SSA) spectrum\footnote{However, the population of the radiating electrons is still under debate \citep{Ho2021,Margalit2021b}.}, which possibly arises when an external shock interacts with a dense circumstellar medium (CSM) of a steep density profile.
	The X-ray emission observed in all FBOTs does not align with the electron population inferred from the SSA radio spectrum. The spectral and temporal evolution, as well as the variability of the X-ray emission in AT2018cow, suggest that the X-ray signal is powered by a rapidly evolving central engine, a magnetar or an accretion disk around a black hole \citep{Ho2019,Margutti2019,Quataert2019,Pasham2021}.
	
	The UV/optical/IR emission reveals some of the most intriguing features of FBOTs:
	i) fast rise ($ \sim $ day) and luminous ($ \sim 10^{44}~{\rm erg~s^{-1}} $) lightcurve;
	ii) quasi-thermal optical spectrum;
	iii) broad hydrogen (H) emission features, hinting at a small amount of ejecta moving at sub- to mildly-relativistic velocities of $ v \gtrsim 0.1 $ c;
	iv) UV/optical/IR luminosity in AT2018lug, AT2018cow, SN2018bcc, AT2020xnd and AT2020mrf rapidly decays with $ L \propto t^{-\alpha} $, where $ \alpha \gtrsim 2 $ \citep{Margutti2019,Ho2020,Karamehmetoglu2021,Perley2021};
	v) plateau of the temperature or even an increase in the effective temperature at late times in AT2018cow, associated with a receding photosphere. Such increase has already been observed previously in events with very high luminosities \citep{Pursiainen2018}.
	The rapid drop observed in the photospheric radius at $ t \sim 20 $ days, as well as the X-ray emission from the central engine, point at a non-negligible degree of asymmetry in the outflow, with multiple emitting components \citep{Margutti2019}.
	
	The above plentiful characteristics of the UV/optical/IR emission challenge models that attempt to interpret the underlying physics of FBOTs.
	A prime candidate for explaining the optical emission is central engine emission reprocessing by the surrounding asymmetric dense CSM \citep{Margutti2019,Piro2020,Uno2020}.
	Physical systems within this scenario include spin-down of millisecond magnetars and tidal disruption events \citep[e.g.,][]{Liu2018,Kuin2019,Perley2019,Kremer2021}, with the latter being disfavored due to the dense CSM.
	%for which the luminosity falls somewhat slower than what FBOTs suggest \citep[][however see \citealt{Metzger2018} for possible reconciliation if fallback of stellar material takes place]{Piro2020}.
	Other proposed scenarios for FBOTs include accretion disks following the collapse of massive star to a BH \citep{Kashiyama2015,Quataert2019}, interaction of pulsational pair-instability SN with the CSM \citep{Leung2020}, electron-capture collapse following the merger of binary white dwarfs \citep{Lyutikov2019}, and common envelope jets SNe \citep[CEJSN;][which we discuss in \S\ref{sec:discussion}]{Soker2019}.
	
	The emergence of FBOTs in star-forming galaxies suggests that they are likely to be associated with the collapse of massive stars. However, the photospheric evolution, peak optical luminosity, fast decay and fast velocities indicate that these are not typical SNe powered by the radioactive decay of expanding photospheric shells.
	The fast rise and fall times, as well as the interaction of the ejecta with the CSM imply that numerous FBOTs could be related to the special stripped envelope SNe \citep[see e.g.,][]{Ho2021,Pellegrino2021}, possibly of type Ib/c, for which asphericity as it is for FBOTs can also be common \citep{Maeda2008}. The progenitors of some of these SNe are considered to be massive Wolf-Rayet (WR) stars that shed their outer H and helium (He) layers through strong winds \citep{Crowther2007}, which may constitute the necessitated dense CSM. Recently, \citet{Maeda2022} recently found that the CSM density distribution in SNe of type Ibn can be $ \rho \propto r^{-3} $, as implied by FBOT radio lightcurves.
	While the spectrum of such SNe (e.g., type Ibn) is characterized by narrow emission lines, in contrast with that found in FBOTs, some of these WR stars could support the launching of relativistic $ \gamma $-ray burst (GRB) jets, and give rise to broad emission lines. Although all detected GRBs to date are associated with SNe of type Ic \citep[e.g.,][]{Cano2017}, it is likely that jets are also launched in WR stars that keep their He envelope (i.e., progenitors of SN Ib), but fail to break free \citep[e.g.,][]{Mazzali2008,Margutti2014,Nakar2015,Sobacchi2017}.
	
	Previous studies have explored the possible existence of GRB jets in FBOTs.
	\citet{Margutti2019} found that the broad spectral lines have velocities comparable with SNe Ic associated with GRBs.
	\citet{Coppejans2020} utilized radio observations of superluminous SNe (SLSNe-I) to find that off-axis or uncollimated jets can be consistent with the data \citep[as we show next, in both cases the emission is powered by the jet's cocoon, see also][]{Gottlieb2018b,Gottlieb2021b}.
	Similarly, \citet{Mooley2021} recently argued that the radio signal of J1419+3940, which resembles FBOTs in many aspects, arises from an off-axis GRB, and \citet{Ho2020} found that the luminous, long-lived radio emission of AT2018lug imitates that of long GRB jets.
	Finally, \citet{Prentice2018,Margutti2019} discussed the possibility that the central engine is a millisecond magnetar and concluded that this would require a strong magnetic field of $ B \approx 10^{15} $~G. Magnetars with such strong fields are considered to generate relativistic jets \citep{Metzger2011}.
	Despite the above independent arguments, which favor the inclusion of GRB jets in FBOT models, such models are yet to be formulated and FBOT-associated jets are yet to be directly detected \citep[e.g.,][]{Bietenholz2020}. If a jet indeed fails to break out from the progenitor star, or beams its emission away from us, what observational signatures would this lead to in such events?
	
	Regardless of the jet fate inside the extended envelope, its interaction with the envelope forms a high-pressure shocked material called the cocoon. It engulfs the jet and may reach relativistic velocities \citep[see e.g.,][]{MacFadyen2001,RamirezRuiz2002,Zhang2003,Lazzati2005}.
	A mildly-relativistic cocoon has been previously suggested as the origin of fast moving material found in observational data of CCSNe that lost either all or most of their H envelope; some harbored GRB jets and some did not \citep{Piran2019}. Specifically, several CCSNe, both of type Ib \citep[e.g., SN2008,][]{Mazzali2008} and Ic \citep[e.g., SN2002ap,][]{Mazzali2002}, revealed fast evolving lightcurves and broad absorption lines in their early (first weeks) spectra, similar to FBOTs \citep[e.g.,][]{Perley2019}.
	Owing to the cocoon opening angle exceeding that of the GRB jets, the expected volumetric rate of cocoon emission is higher than that of GRBs by a factor of a few, and thus similar to that of FBOTs \citep{Coppejans2020,Ho2020}. 
	
	In this paper, we explore the possibility that FBOTs emerge in the aftermath of a massive star collapse. The star could be H-rich or SN Ib progenitor with H-rich CSM if the jet is choked; or SN Ic progenitor if the jet breaks out. In all cases the central engine launches a relativistic jet which is accompanied by a cocoon. As the cocoon pierces the stellar envelope, it expands adiabatically and powers a quasi-thermal cooling emission signal \citep{Nakar2017,Gottlieb2018a,Gottlieb2020a,Suzuki2021}. We motivate the cocoon model in the view of current observational data, because FBOTs in our model:
	\\i) are associated with the deaths of massive stars, as suggested by the host star-forming galaxies, and the somewhat higher event rate of FBOTs compared to GRBs. In our model this is expected from the wide angle cocoon emission.
	\\ii) are the result of an active central engine as suggested by the X-ray emission, which is expected to launch a relativistic jet that forms a cocoon.
	\\iii) yield a variety of high velocities $ v \gtrsim 0.1 $ c \citep{Gottlieb2021a}, and give rise to the observed absorption features \citep{Piran2019}.
	\\iv) anticipated to possess a total energy of $ 10^{50}-10^{51}~{\rm erg} $ \citep[see e.g.,][]{Bromberg2011,Harrison2018}, consistent with that inferred from the UV/optical/IR emission;
	\\v) introduce an asymmetric light ejecta with an angular distribution. Thus, the optical emission in AT2018cow can be explained by the early ($ t \lesssim 20 $ days) cocoon emission, followed by cooling of SN shocked stellar material that moves at $ v \lesssim 0.01 $ c. Such a behavior was previously observed in SN 2017iuk where the cocoon dominated the first epoch of emission, before being outshone by the SN \citep{Izzo2019,Piran2019}.
	\\vi) produce fast decay in luminosity with $ \alpha \gtrsim 2 $, and as we show here is also expected from the cocoon cooling emission.
	\\vii) feature a forward shock that powers synchrotron emission while it propagates in the dense CSM expected around SN Ib progenitors. The radio lightcurves ($ \sim 10~{\rm GHz} $) of AT2018cow and AT2020xnd feature a continuous rise until they peak at $ t \approx 100 $ days. Such a lightcurve naturally emerges due to early cocoon-dominated emission followed by jet emission \citep[e.g.,][]{Gottlieb2019a}. We note that this might also be the case in AT2018lug since the first radio detection was months after the event. By that time, had there been a jet, it might have already decelerated to take the shape of a jet observed on-axis.
	
	\section{Hydrodynamic evolution of the outflow}\label{sec:hydro}
	
	The outflow ejected during the collapse of a massive star can be generated through three different physical processes:
	1. SN shocked matter (SSM): stellar material shocked by the SN shock wave to subrelativistic velocities.
	2. A relativistic jet from the vicinity of the compact object formed at the center of the collapsing star.
	3. Jet shocked matter (cocoon): material shocked by the jet, resulting in a range of velocities, from subrelativistic to mildly-relativistic ones. The interaction of the jet with the stellar envelope leads to the formation of a double shocked layer cocoon: shocked jet material which comprises the inner and faster part of the cocoon, and shocked stellar material in the outer part of the cocoon \citep{Bromberg2011}.
	
	While a jet can either break out from the star or be choked inside, depending on the specific structure of the progenitor star \citep{Gottlieb2021c}, the hot cocoon is likely to ultimately break out from the star, with the fastest cocoon velocity depending on the jet fate.
	If the jet breaks out, the cocoon reaches mildly-relativistic velocities $ \Gamma_c \sim 3 $ \citep{Gottlieb2021a}, whereas if the jet fails, the fastest cocoon velocity primarily depends on the location in the star where it was choked.
	The slowest cocoon velocity is given by $ v_s \approx \sqrt{E_c/M_\star} $, where $ E_c $ is the cocoon energy and $ M_\star $ is the stellar mass \citep{Eisenberg2022}. If the jet launching is accompanied by a SN explosion, it is likely that the SN energy exceeds that of the cocoon $ E_{\rm SN} \gg E_c $. Therefore, the SSM dominates at low velocities, such that the slowest cocoon velocity is $ v_s \approx \sqrt{2E_{\rm SN}/M_\star} $.
	Line profiles in superluminous SNe suggest that the fastest ejecta from the explosion is moving at a few percent the speed of light \citep{Quimby2011,Quimby2018,Nicholl2015,Gal-Yam2019}. For any reasonable jet energies, the cocoon is moving faster than the SSM, thus it will outrun the expanding SSM and dominate the emission at $ v \gtrsim 0.01 $ c.
	
	Numerical simulations suggest that due to the mixing in the cocoon, the cocoon energy is distributed roughly uniformly in the logarithm of the proper-velocity, between $ {\rm log}(v_s/c) $ and $ {\rm log}(\Gamma_c) $ \citep[][]{Gottlieb2021a}.
	If the cocoon is accompanied by an SSM, the energy distribution at $ v \lesssim 0.01 $ c is dictated by the properties of the SSM. The transition from cocoon dominated distribution at high velocity to SSM dominated at low velocities \citep[see e.g. the SN+jet energy distribution in][]{Eisenberg2022} could feature an abrupt change in the emission profile. For example, the rise in the temperature (rapid drop in the photosphere) in AT2018cow may be explained by the sharp fall in the velocity of the dominating emitting component.
	
	\begin{figure*}
		\centering
		\includegraphics[scale=0.53,trim=3 0 0 50]{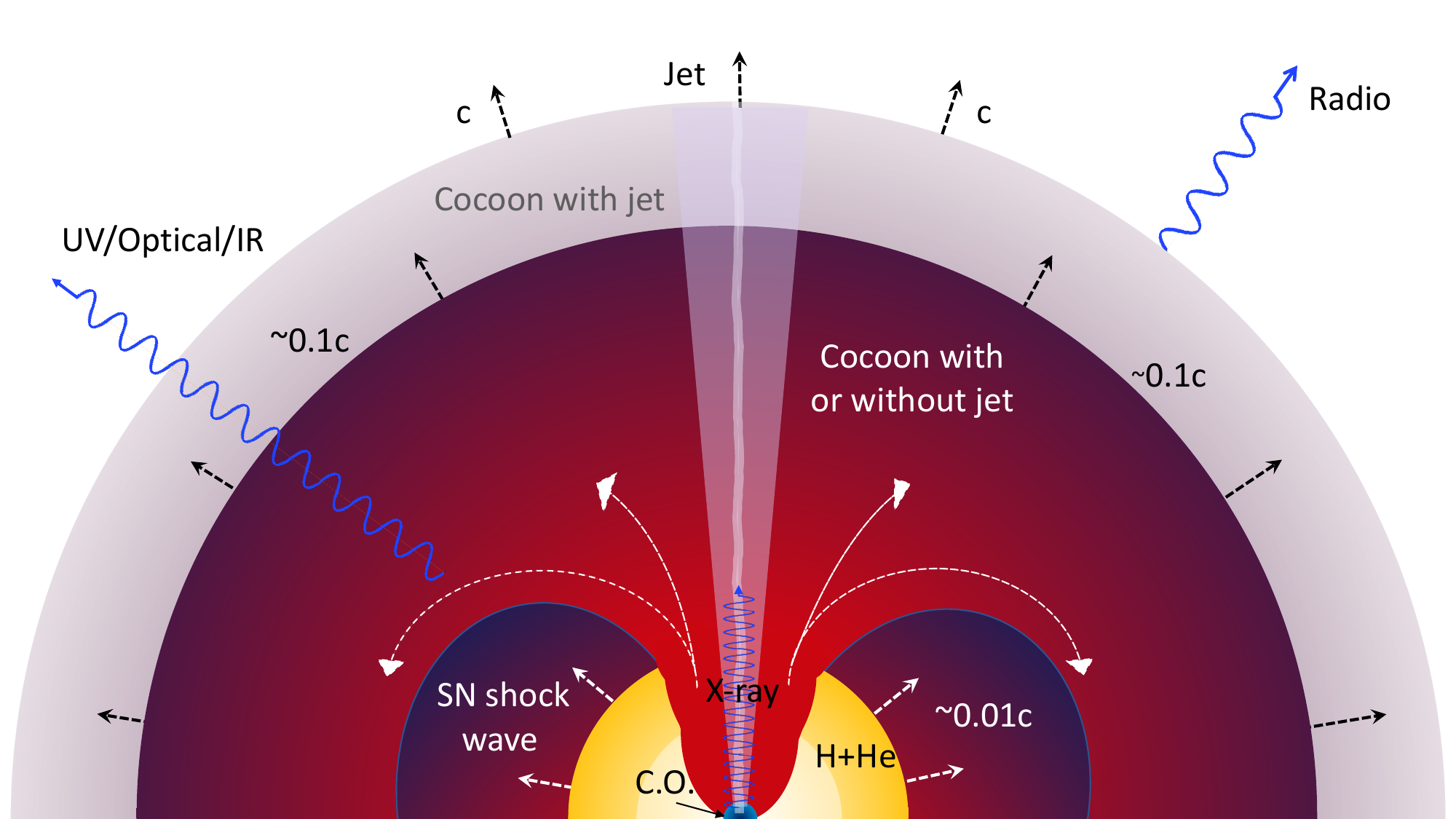}
		\caption[]{
			Schematic illustration of the emerging structure following stellar explosion and launching of a relativistic jet. The jet shocks stellar material, driving a cocoon which is expanding to large angles at $ v_c \gtrsim 0.1 $ c, surpassing the SN shock wave which trails behind at $ v_s \sim 0.01 $ c.
			If the jet manages to drill successfully  through the stellar envelope to break out from the star (yellow), it is accompanied by a cocoon whose fastest velocity in $ \Gamma_c \approx 3 $ (grey area). If however, the jet is suffocated inside the star, the cocoon is slower.
			The relativistic jet and inner cocoon form a low density region close to the polar axis, thereby facilitating the X-rays emitted from the compact object (C.O.; blue) to diffuse out of the thick stellar envelope. The radio emission is generated via synchrotron emission during the external shock-CSM interaction.
		}
		\label{fig:illustration}
	\end{figure*}
	
	Fig. \ref{fig:illustration} illustrates the structure of cocoon+SSM outflow when all components expand homologously. Each element reaches the homologous phase when it approximately arrives at $ \sim 2R_\star $, where $ R_\star $ is the stellar radius.
	The only component whose existence is guaranteed is the cocoon. Its slowest and fastest velocities depend on the SN energy and the jet fate, respectively.
	The grey area, which includes a relativistic cocoon, emerges if the jet, generated by the compact object (C.O.; blue), breaks out from the star (yellow). Otherwise, the asymptotic velocity of the fastest cocoon depends on the location where the jet is choked.
	
	\section{UV/optical/IR light}
	
	We suggest that the UV/optical/IR signals in FBOTs are powered by the cooling emission of the cocoon. If the cocoon is accompanied by SSM, the cocoon emission is followed by cooling of the SN powered stellar envelope which emerges once the cocoon becomes optically thin (Fig. \ref{fig:illustration}). For simplicity we ignore potential effects of a dense CSM on shock cooling \citep[see][]{Margalit2021}.
	
	\subsection{Early times: Cocoon cooling}\label{sec:early_emission}
	
	The cocoon cooling emission naturally explains the broad absorption feature in FBOTs \citep{Piran2019}, and the quasi-thermal spectrum at $ T \sim 10^4 $ K.
	For estimating the peak time of the emission, consider a GRB jet energy with a typical total luminosity $ L \approx 10^{50}~{\rm erg~s^{-1}} $. Observations suggest that a GRB jet spends $ t_b \approx 10 $ s inside the progenitor star before breaking out \citep{Bromberg2012}. During this time it inflates the cocoon that then breaks out with energy $ E_c \approx Lt_b \approx 10^{51}~{\rm erg} $.
	The cooling emission of an element moving at velocity $ v $, is radiated at the trapping radius $ r_t $, where $ \tau(r_t) = c/v $, and
	\begin{equation}\label{eq:tau}
		\tau = \frac{E_c\kappa}{2\pi \beta_c^4t^2c^4}~,
	\end{equation}
	where $ \beta_c $ is the characteristic dimensionless fastest velocity of the outflow, $ \kappa $ is the opacity of the outflow, and we assume homologous expansion.
	That implies that the emission peak time is
	\begin{equation}\label{eq:tp}
		t_p = \left(\frac{E_c\kappa}{2\pi \beta_c^3 c^4}\right)^{1/2} \approx 1.6~(E_{c,51}\beta_{c,0.1}^{-3}\kappa_{0.1})^{0.5}~{\rm days}~,
	\end{equation}
	powered by ejecta mass of
	\begin{equation}
	    M_c \approx 2E_c(\beta_cc)^{-2} = 0.1E_{c,51}\beta_{c,0.1}^{-2} \msun~,
	\end{equation}
	where subscript $ x $ denotes the normalized value in c.g.s. units (units of $ E_c $ are in $ 10^x $).
	While the observed FBOT peak time is also on a $ \sim $~day timescale, the actual peak time is not well constrained as FBOT observations feature a monotonic decay in the light curve from the first data point.
	
	To date, all inferred velocities of FBOTs have been $ v \lesssim 0.6 c $, such that relativistic effects can be neglected to a first approximation. Therefore, although in general the cocoon has both radial and angular distributions \citep{Gottlieb2021a}, the angular dependency does not play an important role in the emission, enabling us to consider only the radial distribution. Over time the emission is generated by deeper layers in the outflow. This implies that the radial (velocity) distribution governs the temporal evolution of the lightcurve.
	
	Consider a cocoon with (radial) energy distribution $ E \propto \beta^{-\epsilon} $. Eq. \ref{eq:tau} dictates that in the subrelativistic regime, the velocity of the emitting layer at $ r_t $ changes with time as\footnote{The opacity may change in time with the temperature, however since the temperature only changes by a factor of $ \sim 2 $, we approximate the opacity to be constant in time.}
	\begin{equation}\label{eq:v}
		\beta \propto t^{-\frac{2}{\epsilon+3}}~.
	\end{equation}
	Assuming homologous expansion, it follows that the internal energy density of a shell moving at $ \beta $ is $ p \propto \beta^{\epsilon+2} $.
	Adiabatic cooling with polytropic index of radiation-dominated gas, $ \gamma = 4/3 $, yields
	\begin{equation}\label{eq:p}
		p(r_t,\beta) \propto t^{-\frac{2\epsilon+4}{3+\epsilon}}~.
	\end{equation}
	The cooling luminosity is obtained by the flux through the photosphere, and from Eqs. \ref{eq:v} and \ref{eq:p} we find \citep[see also][]{Piro2018}
	\begin{equation}\label{eq:L}
		L = 4\pi r_t^2 \beta p(r_t,\beta)c \propto t^{-\frac{4}{\epsilon+3}}~.
	\end{equation}
	
	Plugging $ \epsilon = 0 $, as suggested by numerical studies, into Eq. \ref{eq:L}, we find that $ L \appropto t^{-4/3} $.
	We note that a more accurate treatment would consider only the angular distribution of inner cocoon that dominates the emission. Numerical results show that for the inner cocoon the energy power-law index becomes $ \epsilon \approx -1 $ \citep[see e.g. figure 3 in][]{Eisenberg2022}. This value corresponds to $ L \appropto t^{-2} $, which is more accurately consistent with the evolution of the best studied optical lightcurves, AT2018lug and AT2018cow \citep[e.g.,][who found $ L \propto t^{-2.5} $]{Margutti2019,Perley2019}.

	By equating Eq. \ref{eq:tau} to unity, we find the temporal evolution of the photospheric shell velocity
	\begin{equation}\label{eq:rp}
		\vp \propto t^{-\frac{2}{\epsilon+4}}~,
	\end{equation}
	from which we obtain the effective temperature
	\begin{equation}\label{eq:T}
		T = \left(\frac{L}{4\pi\sigma_{T} (\vp t)^2}\right)^{1/4}\propto t^{-\left(\frac{1}{\epsilon+3}+\frac{1}{\epsilon+4}+\frac{\epsilon}{2\epsilon+8}\right)} \propto t^{-0.6}~,
	\end{equation}
	consistent with AT2018cow observations, where we used again $ \epsilon = -0.5 $.
    
	The aforementioned analysis applies to both jets that successfully break out from the star and those which are choked just before breakout. In both cases the cocoon structure is similar. We note that we can rule out the possibility of an uncollimated outflow powering the emission in FBOTs. Such an outflow cannot produce a jet that successfully breaks out, instead it will produce a quasi-spherical explosion likely deep inside the star. The resulting energy distribution will be similar to that of a SN, i.e., dominated by a radial structure with a distinct peak, and inconsistent with the observed optical lightcurves.
	Another problem with uncollimated outflows is the outflow mass that would be significantly higher than FBOT observations suggest (and the large amount of energy required to accelerate most of the star to the inferred high velocities). For example, in AT2018cow, $ v(t) \propto t^{-\zeta} $ where $ \zeta > 1 $ is increasing with time. Therefore, if the opacity does not change significantly with time, most of the energy is in the fast shells, so the shell at $ v_{\rm max} \approx 0.1 c$ constitutes most of the ejecta energy. Since it is an uncollimated outflow, the required energy is $ E \approx 0.5M_\star v_{\rm max}^2 \approx 10^{53}~{\rm erg} $.
	This energy is very high, had it been injected into a wide angle, we should have seen many more of such events. In particular, it will be essentially indistinguishable from regular SN explosions.
	We conclude that if the jet is choked, it is not because it fails to collimate, but due to its inability to penetrate the thick envelope, presumably owing to the presence of an extended envelope as might be indicated by the broad emission features of H.
	
	\subsection{Late times: SN cooling envelope}\label{sec:late_emission}
	
	If the GRB is accompanied by a SN shock wave, the low velocity distribution reflects the structure of the SSM (\S\ref{sec:hydro}). Once the equatorial parts of the cocoon turn optically thin, the SSM trailing behind becomes visible. The transition from cocoon to SSM dominated emission is reflected in the lightcurve and the spectrum. Specifically, it is anticipated that the photosphere will drop to similar radii and consequently the luminosity and/or the temperature will exhibit a sharp transition as well.
	
	Signs of such a transition might have been detected.
	Unlike most FBOTs, which show no signs of change in the lightcurve, presumably due to observational limitations, AT2018cow featured a clear transition around 20 days. At this time the temperature started to rise, the effective photosphere significantly receded, and H and He lines started to develop around 1000 km/s \citep{Margutti2019,Perley2019}.
	The luminosity at 20 days, $ \sim 5\times 10^{42}~{\rm erg~s^{-1}} $, is consistent with observations of SN II lightcurves. The temperature at $ \sim 20 $ days implies that the emitting shell was moving at $ \sim 2\times 10^3~{\rm km~s^{-1}} $, consistent with our model.
	
	\section{Compatibility with radio and X-ray}
	\label{sec:bands}
	
	\subsection{Radio}
	
	The radio emission in FBOTs has been shown to be consistent with an SSA spectrum powered by the interaction of an external shock with a dense CSM. In our model the external shock is energized by the fast cocoon which expands into the dense CSM surrounding the massive star (Fig. \ref{fig:illustration}).
	The radio lightcurves of AT2018cow and AT2020xnd feature a similar evolution in time with a steady rise on a timescale of $ \sim 100 $ days before a rapid decay. In AT2018lug the first radio detection was months after the explosion when the lightcurve started to fade, possibly preceded by a similar initial rise.
	This behavior is naturally explained by our model, as the cocoon synchrotron emission dominates the early rise before the jet core decelerates enough to enter the observer's line of sight. Once the jet core is visible, the lightcurve is similar to an on-axis GRB afterglow emission where a continuous decay is present.
	The peak time is determined by the jet energy, ambient density and the inclination angle of the system.
	This scenario has already been observed in the broadband afterglow of the neutron star (NS) binary merger GW170817 \citep[see][for reviews]{Nakar2020,Margutti2020}. However, here the rapid decay in the lightcurve suggests that the ambient medium has a steep radial profile, as expected from the winds surrounding the star. We leave a detailed calculation of the radio emission from the cocoon for future work.
	
	Note that even if the jet is choked after crossing a significant part of the star, the outflow distribution has an angular dependency that can yield the turnover on the observed timescales. However, the fast decay of FBOT radio lightcurves favors the presence of a relativistic jet \citep[see e.g.][]{Nakar2018}.
	By the same token, an uncollimated outflow cannot account for the observed radio lightcurves. The reason is twofold. First, such an outflow has a radial structure and thus the decay after the peak is expected to be gradual. Second, a radial outflow with the physical properties of the cocoon would peak only years after the explosion \citep[see e.g.][]{Piran2013,Kathirgamaraju2019}. This conclusion complements the arguments against uncollimated outflows presented in \S\ref{sec:early_emission} and \citet{Coppejans2018}.
	
	\subsection{X-rays}
	
	Observations during the first $ \sim $ hour after the explosion may reveal free--free emission in soft X-rays, powered by the cocoon-SSM interaction. However, to date all X-ray observations associated with FBOTs are on $ \sim $ day timescales. Following \citet{Margutti2019}, we consider the central engine as the source of the variable X-ray emission. The central engines can be millisecond magnetars that are known to power relativistic jets \citep{Metzger2011}, or fluctuating BH accretion disks \citep{Quataert2019}, which recently were found to naturally emerge in collapsars, and produce GRB jets \citep{Gottlieb2021c}.
	In our model, the jet propagates in the star and forms a low density funnel. After it breaks out, the funnel is refilled by the advection of cocoon material. As the outflow expands, the funnel becomes optically thin to X-ray photons which can then diffuse out to the observer through the funnel (Fig. \ref{fig:illustration}).
	
	We estimate whether X-rays generated by the compact object can reach the observer by the time of X-ray observations on a $ \sim $ day timescale. The optical depth from the compact object to the observer is given by
	\begin{equation}\label{eq:taux}
		\tau(r) = \int_{{\rm C.O.}}^\infty{\rho(r)\kappa\Gamma(r)[1-\beta(r)]} dr~,
	\end{equation}
	where $ \rho $ and $ \Gamma $ are the mass density and Lorentz factor, respectively. The highly ionized matter in small latitudes is dominated by electron scattering, $ \kappa \approx 0.2~{\rm cm^2 g^{-1}} $, where in the observed X-ray spectrum Klein-Nishina corrections are negligible.
	
	The jet breaks out from the star with a mildly-relativistic head \citep{Gottlieb2019b}, and starts accelerating to its asymptotic Lorentz factor of a few hundreds. Numerical simulations show that upon breakout, the jets' magnetic field is subdominant such that the jet can be treated as hydrodynamic \citep{Gottlieb2021c}. Since hydrodynamic jets accelerate as $ \Gamma \propto r $, they spend $ \sim 2 $ orders magnitude in time in the acceleration phase. During this time the jet mass density evolves as $ \rho(r) \propto r^{-3} $.
	Plugging into Eq. \ref{eq:taux}, we find that during the jet acceleration the optical depth evolves as $ \tau \propto t^{-3} $, thus the optical depth along the jet drops by a factor $ \sim 10^6 $ before the jet reaches its asymptotic velocity.
	Analytic and numerical estimates show that the optical depth behind the collimation shock at the jet base is $ \tau \approx 3 \times 10^5 $ at the time of breakout \citep{Gottlieb2019b}. That implies that hydrodynamic\footnote{Magnetized jets are less polluted with baryons \citep{Gottlieb2020b}, and therefore maintain an even lower optically depth.} jets become optically thin in less than $ \sim $ hour.
	
	Next we consider the expansion of the light shocked stellar material that fills the funnel after the jet breakout. This is also the relevant scenario when the jet fails to break out. The optical depth in this regime depends on the location at which the jet failed to break out, i.e. the more the jet advanced in the star, the lower the optical depth. Our simulation (\S\ref{sec:numerics}) shows that the optical depth on the axis is $ \sim 10^5 $ at the homologous phase of the cocoon ($ \sim 10^3 $ s). Homologous expansion of the cocoon (Eq. \ref{eq:tau}) implies that $ \tau \propto t^{-2} $. Therefore, the X-ray photons can diffuse out from the central engine close to the jet axis after $ \sim $ 3 days. This timescale is consistent with the early X-ray observations of FBOTs.
	
	\section{Numerical results}\label{sec:numerics}
	
	We demonstrate the compatibility of the cocoon cooling emission with the UV/optical/IR lightcurves of FBOTs by carrying out a 2D\footnote{We emphasize that even though 2D axisymmetric simulations have been shown to be inaccurate when addressing the mildly-relativistic and relativistic components of the outflow \citep{Gottlieb2021a}, our interest here is in the subrelativistic outflow distribution, which was found to be in good agreement with 3D models \citep{Gottlieb2018a,Eisenberg2022}.} relativistic hydrodynamic simulation of a jet launched into a stellar envelope, using the code \textsc{pluto} \citep{Mignone2007}. We post-process the simulation output semi-analytically to produce estimates of the cooling emission lightcurve and temperature.
	We compare the lightcurves with the observational data of AT2018cow, since this is the best studied FBOT to date.
	
	\subsection{Simulation setup}
	
	We inject into a stellar envelope a relativistic (initial Lorentz factor $ \Gamma_0 = 10 $), narrow (initial opening angle $ \theta_0 = 4^\circ $), hot jet with specific enthalpy $ h_0 \equiv 1+4p_0/\rho_0 = 100 $, where $ p_0 $ and $ \rho_0 $ are initial pressure and density, respectively. The jet operates for 100 seconds with a total (two-sided) luminosity, $ L = 4\times 10^{49}~{\rm erg~s^{-1}} $.
	We choose the mass and radius of the \emph{static, post-stripped} stellar envelope to be $ M_\star = 4 \msun $ and $ R_\star = 6\times 10^{11} $ cm, respectively; the stellar density profile is $ \rho(r) = \rho_0(1-r/R_\star)^3 $, where $ \rho_0 $ is set by the stellar mass.
	The jet cylindrical radius of injection is $ r_0 = z_0\theta_0 $, where $ z_0 = 0.01R_\star $ is the height of the lower boundary at which we inject the jet.
	
	We use an ideal gas equation of state with a polytropic index of 4/3, as appropriate for radiation-dominated gas. This equation of state applies at all times of interest since the post-processing emission estimates take place well before the outflow becomes gas pressure dominated.
	Our 2D Cylindrical grid includes a uniform cell distribution followed by a logarithmic one, on each axis. On the $ \hat{r} $-axis, the inner $ 0.01R_\star $ includes a uniform distribution of 100 cells, followed by additional logarithmically spaced 1200 cells out to $ 100R_\star $. On the $ \hat{z} $-axis, 800 uniform cells span from $ z_0 = 0.01R_\star $ to $ R_\star $, followed by a logarithmically spaced 1800 cells out to $ 100R_\star $. The integration is performed with Harten-Lax-van Leer (HLL) solver, Runge-Kutta time stepping, and piecewise parabolic interpolation.
	
\begin{figure}
		\centering
		\includegraphics[scale=0.24,trim=80 0 0 0]{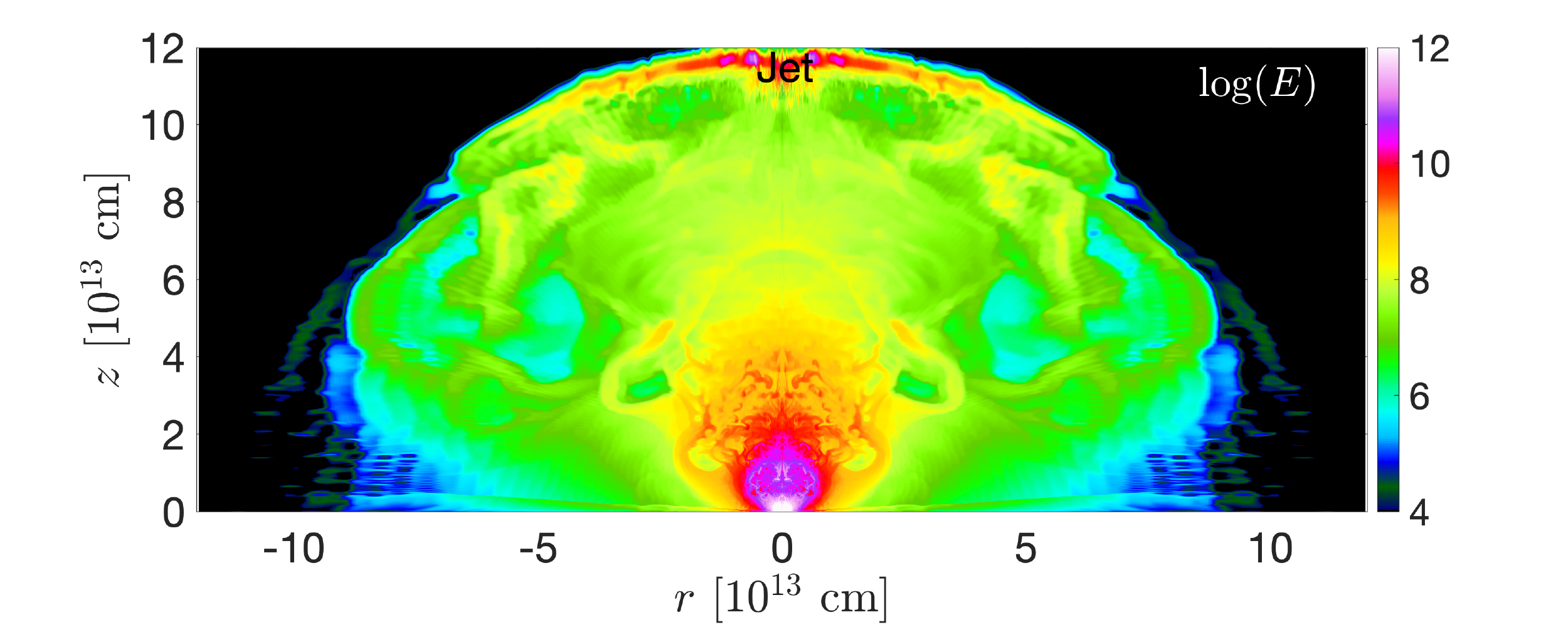}
		\includegraphics[scale=0.22,trim=0 0 0 0]{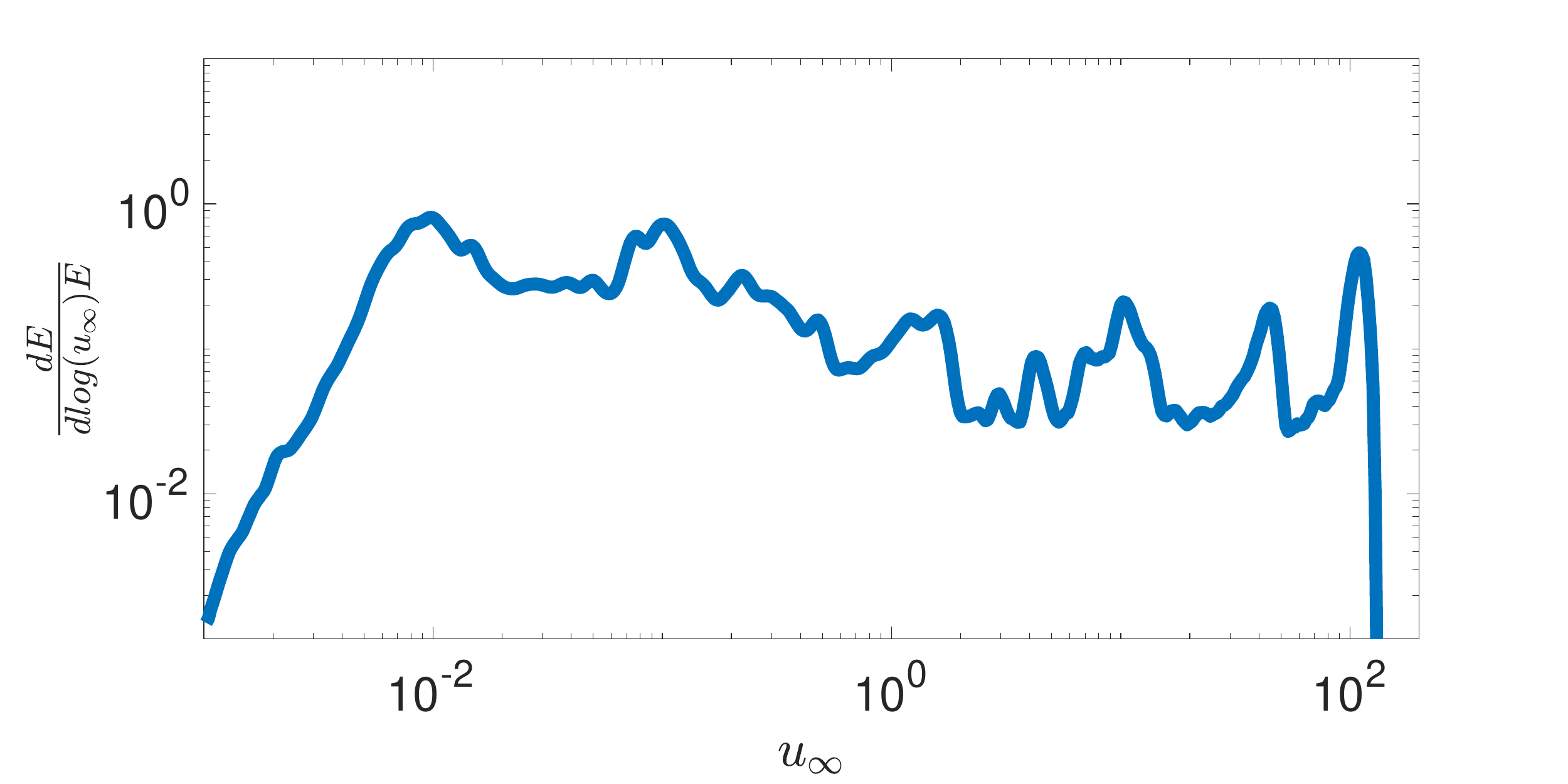}
		\caption[]{
		Following the jet explosion, the emerging outflow (top) yields a distribution of equal energy per decade in asymptotic proper velocity $ u_\infty $ (bottom), as needed for explaining the time evolution of the lightcurve.
		The 2D map shows the logarithm of the kinetic+thermal energy density ($ {\rm erg~cm^{-3}} $) of the outflow accompanies a successful jet (red), 4000 seconds after the explosion.
		The uniform energy distribution in the cocoon (the observed velocities in AT2018cow are $ v \lesssim 0.1 $ c) sets the emission in the first few weeks, and in most cases is insensitive to the jet fate. 
		The distribution is normalized by the total energy of the outflow.
		The cocoon carries $ 0.02\msun $ at $ v > 0.1 $ c, and $ 0.3\msun $ at $ v > 0.02~{\rm c} \approx v_s $.
		}
		\label{fig:simulation}
	\end{figure}

	The top panel of Fig. \ref{fig:simulation} depicts the energy density map 4000 seconds after the jet launch. At this time, the entire star has been shocked and shells moving at $ v \gtrsim 0.01 $ c are homologous. The energetic relativistic jet is seen on the axis in the front in red, and the asymmetric cocoon occupies the remaining outflow.
	The bottom panel displays the energy distribution as a function of the asymptotic proper-velocity of the outflow $ u_\infty $, featuring a quasi-flat distribution at the relevant velocities of the cocoon ($ v \lesssim 0.1 $ c in AT2018cow), as was found by previous numerical studies.
	Note that in general the jet can be choked before breaking out from the star. Although such a scenario entails substantial differences in the relativistic components, it will not affect the energy distribution at $ v \lesssim v_c $, which remains flat \citep{Eisenberg2022}. This implies that the resulting lightcurves are unaffected by the jet fate, as long as the cocoon maximal energy $ v_c $ (which depends on the location of the jet head when it chokes) is higher than the observed maximal velocities.
	
	\subsection{Post-process emission}
	
	\begin{figure}
		\centering
		\includegraphics[scale=0.25,trim=0 0 0 0]{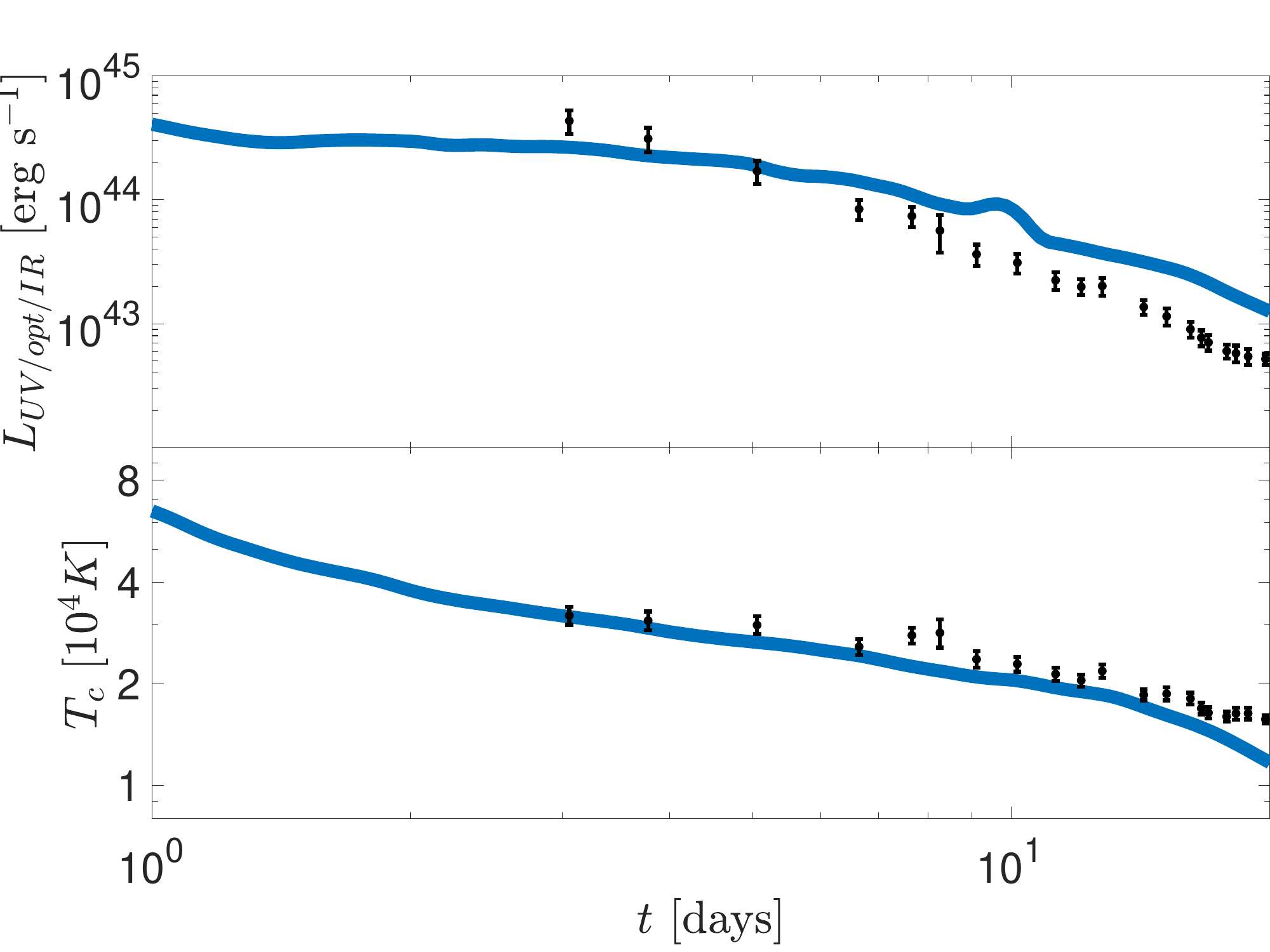}
		\caption[]{
			The cocoon model exhibits an overall trend which is qualitatively similar to the early optical emission of AT2018cow. Shown are the UV/optical/IR luminosity (top panel) and color temperature (bottom panel) of the cocoon cooling emission, against the observed data \citep[black bars;][]{Margutti2019}.
		}
		\label{fig:light_curves}
	\end{figure}
	
	We run the simulation until all shells that move at $ v \gtrsim 0.01 $ c reach homologous expansion. Then, using adiabatic relations, we solve analytically the hydrodynamic evolution of the outflow at later times. We compute the lightcurves semi-analytically as follows. We first calculate the optical depth to infinity at each angle and time, to find the trapping and photospheric radii at which the luminosity and color temperature are set, respectively. Then, using Eq. \ref{eq:L} we calculate the luminosity locally at each angle and time, and integrate it over the entire outflow to obtain the total luminosity. Since the shells of interest are subrelativistic, we do not find significant differences between observers.
	We estimate the temperature by calculating it locally at each angle and time, assuming blackbody emission, and integrating over the solid angle to find the observed spectral luminosity at each frequency and time. Finally, we obtain the color temperature by the frequency for which the spectral luminosity is the highest at each time.
	
	We do not model numerically the SSM emission to account for the late evolution in AT2018cow for multiple reasons:
	first, the simulation is purely hydrodynamic such that it does not include nuclear processes that continuously heat the SSM, and control its dynamics and emission.
	Second, \textsc{pluto} is not a GRMHD code, and has no self-gravity which becomes important at SSM velocities that are comparable to the escape velocity from the stellar surface, $ v \sim 0.01 $ c.
	Last, AT2018cow is the only FBOT for which we were able to confidently constrain the time evolution of the temperature which shows a significant rise. Before additional high quality observations can verify that such behavior is common among FBOTs, the role of the SSM emission remains uncertain.
	
	Fig. \ref{fig:light_curves} depicts the numerical lightcurve as observed by a typical observer, close to the equatorial plane. The UV/optical/IR luminosity (top) and color temperature (bottom) span the range from 1 to 20 days, during which the emission is dominated by UV/optical/IR photons emitted by the cocoon.
	The cocoon emission is consistent with the early optical emission in AT2018cow by up to a factor of two. The lightcurve is derived from the universal flat energy distribution of cocoons, which is independent of the jet fate, as long as it is not choked far from the stellar edge. We conclude that power-law UV/optical/IR lightcurves in FBOTs, including AT2018cow, are generic and do not require fine-tuning of our model.

	\section{Discussion}\label{sec:discussion}
	
	The increased broadband detection rate of FBOTs in star-forming galaxies attracts interest from both observational and theoretical perspectives. While the radio observations seem to be consistent with synchrotron self-absorption, and the variable X-ray signal can be explained by central engine activity, the origin of the fast rise and fall of the optical lightcurve remains poorly understood.
	In this paper, we propose that the fast ($ v \gtrsim 0.1 $ c) component that dominates the early emission is naturally explained by the cocoon, which is made up of the shocked jet and stellar material, and naturally expected as long as a jet is launched from the compact object during the death of a massive star. Our model is motivated by multiple observational arguments:
	\\1. FBOTs feature several observational characteristics that are similar to GRB systems, including: dense CSM owing to strong winds from the star, fast-fading radio emission, and comparable event rates. Owing to the cocoon opening angle, detectable emission should be in similar volumetric rates to FBOTs, somewhat higher than those of GRBs.
	\\2. The central engines in the star can be millisecond magnetars or fluctuating BH accretion disks. In both cases the launch of a relativistic jet and the subsequent formation of a cocoon are expected.
	\\3. The cocoon, which spans a wide velocity range, is expected to have the mildly-relativistic velocities in the range $ 0.1 \lesssim v_c/c \lesssim 0.8 $ to explain the broad absorption features.
	\\4. The cocoon energy and its radial distribution predict early optical emission with the magnitude and fast decay similar to those observed in FBOTs. This emission outshines that of the SN during the first couple of weeks.
	\\5. The inherent aspherical structure of the cocoon, as well as its slower SN shocked material companion, may give rise to observed changes in the lightcurve after a few weeks.
	\\6. The radio emission from the cocoon will show a continuous rise followed by a rapid decay once the jet enters the observer's line of sight. The peak time depends on the jet and CSM properties as well as the viewing angle, and can vary between days to years. A detailed calculation of the radio lightcurve is left for future work.

	Our model is insensitive to whether the jet, which is launched by the central engine, successfully breaks out from or suffocates close to the stellar edge (but rules out uncollimated jets).
	The reason is that the cocoon is primarily constructed during the jet propagation inside the star, so a jet that is choked just before breakout would yield a very similar cocoon to that whose jet successfully breaks out. However, jet emission itself may help to reveal the jet fate. Specifically, the observed rapid decay in the radio lightcurve may imply the presence of a relativistic jet. On the other hand, GRB jets are yet to be detected in systems with H lines, as found in FBOTs, implying that extended H envelopes could choke such jets, leaving the cocoon as the only observational manifestation of a collimated jet.
	
	We emphasize that the progenitors of jets in general do not have to be the same as those of GRBs. For example, if the jet is choked, the star can be a SN II progenitor with H and He shells; or it can be a SN Ib progenitor which is surrounded by a dense H-rich CSM. In the latter case the H lines emerge from the interaction between the outflow and the H-rich CSM, as been found for H-poor SN outflows \citep[e.g.,][]{Chugai2006,Margutti2017,Yan2017,Mauerhan2018}.
	Since the central engine is disconnected from the outer shells, the jet launching is anticipated to be independent of the H and He shells. However, the ability of the jet to break through the star and generate the GRB largely depends on the stellar structure. If the H and He shells were shed prior to the jet launching (SN Ic progenitor), it can produce a GRB and the H lines emerge from the interaction of the outflow with the CSM. Ejection of at least the H shell might be preferred in the context of FBOTs. First, the dense CSM implies that that some non-negligible amount of mass was ejected from the outermost layers of the star (i.e. H). Second, if the presence of a He shell is enough to choke a jet and explain the absence of SN Ib-GRB coincident detections, then having another H shell implies that the jet is choked far from the stellar surface. Jets that are choked deep inside the star can only produce Newtonian cocoons, in some tension with the observed high velocities of FBOTs.
	
	Another FBOT model that involves jets was proposed by \citet{Soker2019}. They considered a CEJSN event, in which a giant star merges with a NS. In their model, the stellar envelope mass is sufficiently low to enable the NS to clear the polar axis before entering the core and launching subrelativistic jets. The relatively baryon-free polar axis allows the jet to propagate freely and power the early emission, followed by late emission from the equatorial outflow. 
	The main difference between the models is the type of explosion and the emission source. While we specifically propose the cocoon emission from CCSNe to power FBOTs, their model considers the broad contribution of subrelativistic jet emission in CEJSN events. The binary system is motivated by the aspherical outflow implied by AT2018cow, however in our model the asphericity is naturally induced by the directionality and collimated nature of the jet.
    We also note that our model is complete in the sense that we provide both a hydrodynamic solution and emission estimates at each frequency. The CEJSN model on the other hand does not specify the emission mechanism that comes into play in each band and time, therefore it remains unclear if the CEJSN model is compatible with the observed AT2018cow multi-band lightcurve. Additionally, a hydrodynamic solution of the CEJSN is essential for ensuring a self-consistent evolution and assessing the role of the jet, cocoon and the funnel.
    %1. It is expected that the low density funnel forms by the passage of the NS will be filled up by advection of baryons from the envelope before the subrelativistic jet reaches it (as we discussed in \S\ref{sec:bands}). This raises the question of whether the loading can remain low to allow the proposed scenario.
    %2. Can a NS launched jet with luminosity $ 10^{47}~{\rm erg~s^{-1}} $ undergo enough mixing in such low density funnel to become asymptotically Newtonian?
    %3. What are the properties of the emerging cocoon in such CEJSN, and what would be its observational role?
    
	Early ($ \sim $ day) detections are key to deciphering the underlying physics of these events towards a better understanding of their emission mechanism. If the central engine is the source of the X-ray emission, then an early X-ray detection may place constraints on the outflow opacity and mass, and may provide clues about the X-ray source and whether it is embedded in a stellar envelope. X-ray detection in the first $ \sim $ hour may also feature free--free emission from the interaction of the outflow with the dense CSM.
	Fast optical detection during the first $ \sim $ hour, as Rubin Observatory may attain, could probe the optical rise, for which spectra may unveil whether relativistic components (such as the inner cocoon) are also present.
	Finally, harder emission in $ \gamma $-rays to UV, depending on the jet energy and viewing angle, is also expected from the cocoon shock breakout, if the viewing angle is not too far from the axis of symmetry (roughly $ 2\Gamma_c^{-1} \approx 40^\circ $). While the specific shape and time of relativistic shock breakout emission are yet to be calculated, and largely depend on the unknown CSM profile, it has certain characteristics \citep[such as its spectrum, see e.g.][]{Nakar2012}. Therefore, a rapid detection of such emission could be essential to verifying the model presented here.
	
	\section*{Acknowledgements}
	
	We thank Noam Soker for helpful comments.
	OG is supported by a CIERA Postdoctoral Fellowship. AT was supported by the NSF awards AST-1815304 and AST-2107839.
	The authors acknowledge the Texas Advanced Computing Center (TACC) at The University of Texas at Austin for providing HPC and visualization resources that have contributed to the research results reported within this paper via the LRAC allocation AST20011 (\url{http://www.tacc.utexas.edu}).
	
	\section*{Data Availability}
	
	The data underlying this article will be shared on reasonable request to the corresponding author.
	
	\bibliographystyle{mnras}
	\bibliography{ref}
	
	\label{lastpage}
\end{document}